\documentclass[%
aps,
reprint,
superscriptaddress,
amsmath,amssymb,
prl,
floatfix,
]{revtex4-2}

\usepackage[utf8]{inputenc} 
\usepackage[T1]{fontenc}

\usepackage{graphicx}
\usepackage{dcolumn}
\usepackage{bm}
\usepackage[colorlinks,urlcolor=Blue,linkcolor=Blue,citecolor=Blue]{hyperref}

\usepackage{textcomp}

\mathchardef\mhyphen="2D

\usepackage{CJK}
\usepackage{upgreek}

\usepackage[dvipsnames]{xcolor}

\usepackage[
top=0.70in, left=0.75in, right=0.75in, bottom=0.82in,
]{geometry}

\begin{document}

\title{Large Momentum Transfer Clock Atom Interferometry\texorpdfstring{\\}{ } on the 689\texorpdfstring{~}{ }nm Intercombination Line of Strontium}%

\begin{CJK*}{UTF8}{gbsn}

\author{Jan Rudolph}
\email[]{jan.rudolph@stanford.edu}
\affiliation{
  Department of Physics, Stanford University, Stanford, California 94305, USA
}%

\author{Thomas Wilkason}
\affiliation{
  Department of Physics, Stanford University, Stanford, California 94305, USA
}%

\author{Megan Nantel}
\affiliation{
Department of Applied Physics, Stanford University, Stanford, California 94305, USA
}%

\author{Hunter Swan}
\affiliation{
  Department of Physics, Stanford University, Stanford, California 94305, USA
}%

\author{Connor M.\ Holland}
\affiliation{
  Department of Physics, Stanford University, Stanford, California 94305, USA
}%

\author{Yijun Jiang (姜一君)}
\affiliation{
Department of Applied Physics, Stanford University, Stanford, California 94305, USA
}%

\author{Benjamin E.\ Garber}
\affiliation{
  Department of Physics, Stanford University, Stanford, California 94305, USA
}%

\author{Samuel P.\ Carman}
\affiliation{
  Department of Physics, Stanford University, Stanford, California 94305, USA
}%

\author{Jason M.\ Hogan}
\email[]{hogan@stanford.edu}
\affiliation{
  Department of Physics, Stanford University, Stanford, California 94305, USA
}%

\date{\today}

\begin{abstract}
We report the first realization of large momentum transfer (LMT) clock atom interferometry. Using single-photon interactions on the strontium ${}^1S_0~ \mhyphen {}^3P_1$ transition, we demonstrate Mach-Zehnder interferometers with state-of-the-art momentum separation of up to $141~\hbar k$ and gradiometers of up to $81~\hbar k$. Moreover, we circumvent excited state decay limitations and extend the gradiometer duration to 50 times the excited state lifetime. Because of the broad velocity acceptance of the interferometry pulses, all experiments are performed with laser-cooled atoms at a temperature of $3~\upmu \text{K}$. This work has applications in high-precision inertial sensing and paves the way for LMT-enhanced clock atom interferometry on even narrower transitions, a key ingredient in proposals for gravitational wave detection and dark matter searches.
\end{abstract}

\maketitle

\end{CJK*}
Atom interferometry (AI) is a versatile and powerful tool in inertial sensing~\cite{Cronin2009,Peters2001,McGuirk2002,Durfee2006} and precision measurements~\cite{Parker2018,Fixler2007,Rosi2014}. Light-pulse atom interferometers split, redirect, and recombine matter waves by imparting photon momenta~\cite{Kasevich1991,Giltner1995}. Their sensitivity to inertial forces can be improved with large momentum transfer (LMT) techniques that use additional light pulses to increase the space-time area of the interferometer~\cite{McGuirk2000}. 
Conventional light-pulse atom interferometry uses two-photon interactions, implemented by a pair of laser beams far detuned from a strong optical line. However, some of the most demanding applications, such as ultralight dark matter searches~\cite{Arvanitaki2018,Bertoldi2019} and gravitational wave detection~\cite{Dimopoulos2008,Yu2011,Chaibi2016,Loriani2019,Zhan2019,Schubert2019}, can benefit from the use of single-photon transitions like the ultranarrow lines typically employed in optical lattice clocks~\cite{Ushijima2015,Bloom2014,Marti2018}. LMT-enhanced clock atom interferometry, based on a sequence of single-photon transitions, was recently proposed~\cite{Graham2013} as a method to reach the required sensitivity while retaining the necessary level of laser noise suppression~\cite{Graham2017}. Proof-of-principle clock atom interferometry without enhanced momentum separation has been performed on the ${}^1S_0~ \mhyphen {}^3P_0$ strontium clock transition~\cite{Hu2017,Hu2019}. Here we demonstrate the first realization of LMT-enhanced clock atom interferometry using the ${}^1S_0~ \mhyphen {}^3P_1$ intercombination line in \textsuperscript{88}Sr.

State-of-the-art LMT atom interferometers employ Raman transitions~\cite{Dickerson2014,Kotru2015}, Bragg transitions~\cite{Muller2008,Ahlers2016}, and Bloch oscillations in optical lattices~\cite{Clade2009,Mueller2009,McDonald2013,Gebbe2019} to scale up the momentum transfer. While optical lattices can generate large momentum separation, the confining potential can cause unwanted systematic effects~\cite{Gochnauer2019,Pagel2019}. In free-space atom interferometry, a total momentum transfer of $112~\hbar k$ has recently been achieved via sequential Bragg transitions in \textsuperscript{174}Yb~\cite{Plotkin-Swing2018}, improving upon the previous record of $102~\hbar k$ in \textsuperscript{87}Rb~\cite{Chiow2011}. One important constraint on further improvements of two-photon LMT techniques is spontaneous emission loss via the short-lived excited state, requiring sizable detunings and laser intensities~\cite{Kovachy2015}.

In contrast to conventional Raman or Bragg atom optics, clock atom interferometry uses narrow-linewidth transitions that are driven resonantly by a single laser beam. For clock transitions to metastable states such as the strontium ${}^1S_0~ \mhyphen {}^3P_0$ transition, the spontaneous emission loss from excited state decay can be negligible due to the $150~\text{s}$ lifetime~\cite{Santra2004}. Furthermore, for this transition spontaneous scattering from other off-resonant lines is suppressed by terahertz detunings such that it can in principle support many thousands of consecutive pulses. However, for efficient interferometry pulses the laser must be frequency stabilized to an optical cavity to reduce its linewidth below the target Rabi frequency.

We show that moderately narrow transitions such as the $7.4~\text{kHz}$ strontium ${}^1S_0~ \mhyphen {}^3P_1$ intercombination line support single-photon Rabi frequencies in the megahertz range, easing the technical requirements for laser linewidth reduction. This high Rabi frequency stems in part from the fact that the single-photon Rabi frequency $\Omega$ is much larger than the two-photon Rabi frequency for the same transition, which is suppressed by a factor of $\Omega/\Delta\! \ll\! 1$, where $\Delta$ is the detuning from the excited state. The short pulse durations we achieve with this transition allow for hundreds of consecutive pulses despite the $21.6~\upmu\text{s}$ excited state lifetime. Additionally, the high pulse bandwidth renders the transitions insensitive to Doppler detunings, for example from velocity offsets due to gravity or from momentum separation between the interferometer arms. While previous two-photon interferometers with state-of-the-art momentum transfer have generally relied on ultracold and quantum degenerate atoms~\cite{Szigeti2012}, we show that this new type of interferometer accepts microkelvin temperature ensembles from a magneto-optical trap (MOT) without reducing the pulse efficiency. Since no additional cooling or velocity selection techniques are required, this has the potential to increase the number of atoms in LMT interferometers by orders of magnitude, further improving the sensitivity of shot-noise-limited sensors.

\begin{figure}[t]
        \includegraphics[width=\linewidth]{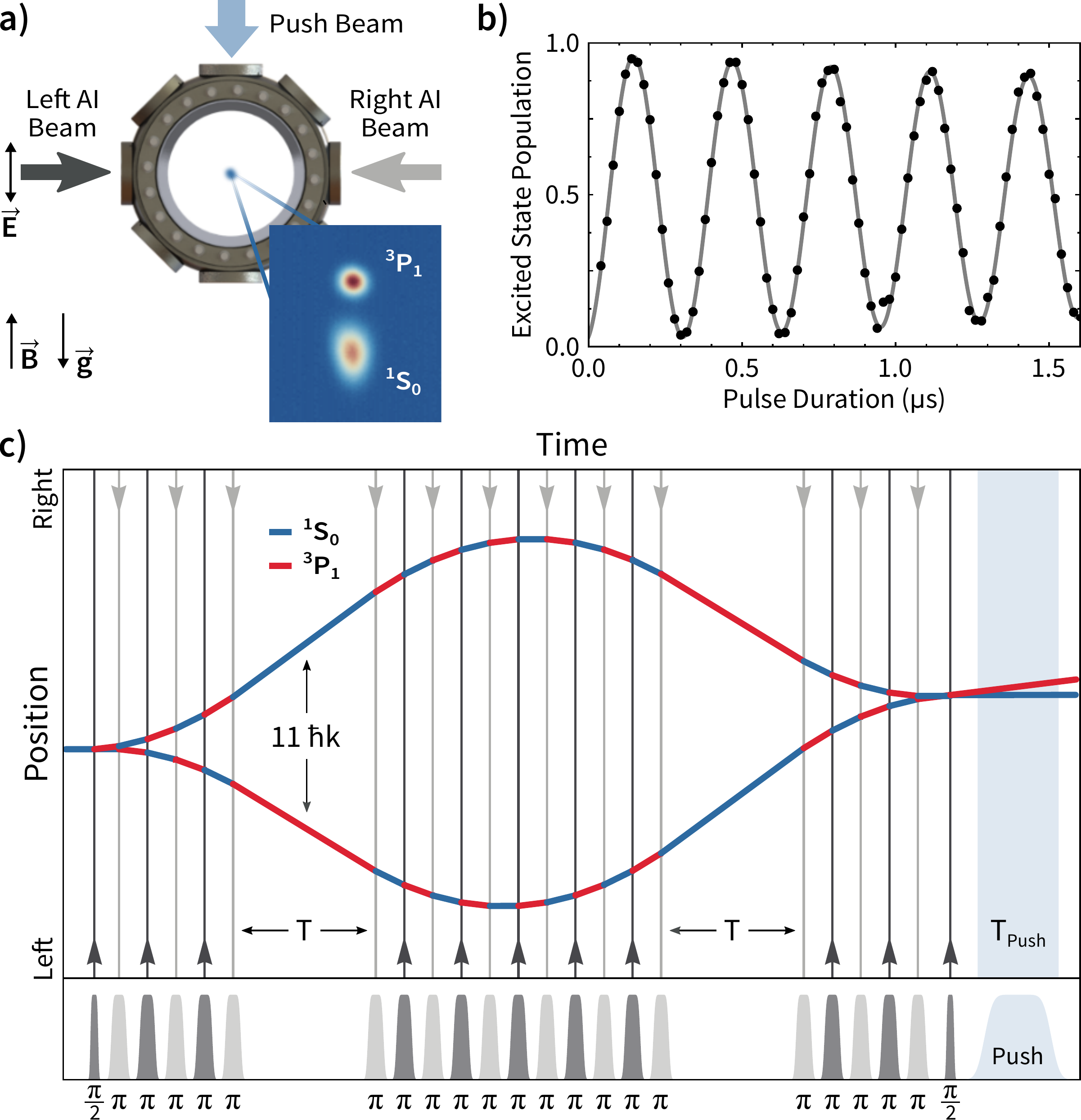}%
     \caption{a) Experimental setup. Two independent, horizontal interferometry beams (dark gray and light gray) interact with the atoms one at a time, from alternating directions. The beams are linearly polarized parallel to the applied magnetic bias field $\roarrow{B}$. A vertical push beam (blue) separates the atomic states for fluorescence imaging (inset). b) A typical Rabi oscillation of the normalized ${}^3P_1$ excited state population, with a Rabi frequency of $3.11~\text{MHz}$ and a $\pi$-pulse efficiency of $98.9\%$. c) Example LMT interferometer space-time diagram (top) and associated pulse sequence (bottom). The alternating pulses from the left beam (dark gray) and the right beam (light gray) interact with both arms of the interferometer, transferring momentum and toggling the atom between the ground (blue) and excited (red) states. The pulses are distributed over three zones, separated by the interrogation time $T$, in between which the direction of momentum transfer is reversed. A push pulse occurs at the end of the sequence.}
    \label{fig:1}
\end{figure}

We prepare typical samples of $10^7$ strontium atoms via a dual-stage MOT on the blue $461~\text{nm}$ transition, followed by the red $689~\text{nm}$ transition~\cite{Katori1999}. The red MOT light is generated by an external cavity diode laser that is cavity stabilized to a linewidth of $\approx 1~\text{kHz}$. Two independent interferometry beams are derived from this same laser source, which is amplified by a tapered amplifier to allow for $100~\text{mW}$ of optical power per beam. Each beam is focused through a pinhole to clean its spatial mode and has a $1.5~\text{mm}$ waist at the location of the atoms, approximately 10 times larger than the typical rms radius of the atom ensembles of $160~\upmu\text{m}$. Polarizers ensure that the polarization of the horizontal interferometry beams is aligned parallel to a vertical magnetic bias field to resonantly drive ${}^1S_0~ \mhyphen {}^3P_1~(m\! =\! 0)$ transitions (see Fig.~\hyperref[fig:1]{1a}). We use a bias field amplitude of $10~\text{G}$ to further suppress unwanted excitations to $m\!=\! \pm 1$ Zeeman sublevels.\smallskip

The interferometry pulse shapes and amplitudes are produced by a nanosecond programmable pulse generator, fast RF switches, and independent single-pass acousto-optical modulators (AOMs) for each of the two beams. We achieve a typical $\pi$-pulse duration of $161~\text{ns}$ and Rabi frequency of $3.11~\text{MHz}$. Thanks to the high pulse bandwidth, the laser frequency is held constant throughout the interferometry sequence, even as the atom velocity changes. Despite the significant rms Doppler width of $24~\text{kHz}$ at an ensemble temperature of $3~\upmu\text{K}$, we reach $\pi$-pulse efficiencies of $(98.9 \pm 0.2)\%$, inferred from the exponential decay of the Rabi oscillation amplitude (see Fig.~\hyperref[fig:1]{1b}). Note that the observed peak normalized excited state population is reduced because some excited state atoms decay during the $T_\text{push}=2~\upmu\text{s}$ push pulse at the end of the sequence. This pulse, from a laser beam resonant with the $461~\text{nm}$ transition, imparts momentum to atoms in the ground state, which leads to a vertical separation of the states after $5~\text{ms}$ time of flight prior to detection. By the time the atoms are illuminated for fluorescence imaging on that same transition, all excited state atoms have decayed to the ground state. Images are formed on a CMOS camera using a 1~:~1 imaging system in the horizontal plane, at an angle of {$45$\textdegree} relative to the interferometry beams.\smallskip

\begin{figure*}[t]
        \includegraphics[width=\linewidth]{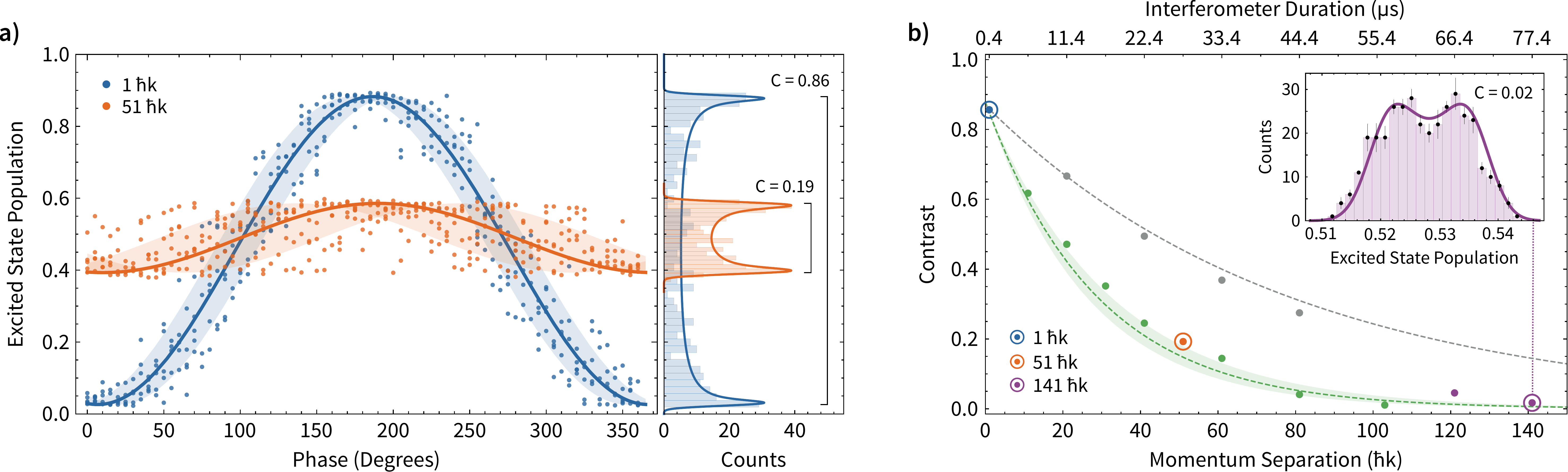}%
     \caption{a) Normalized excited state population of the interferometer versus the interferometer phase, for a momentum separation of $1~\hbar k$ (blue) and $51~\hbar k$ (orange). The contrast $C$ is determined by fitting the population histogram (right), while the phase noise is estimated from the uncertainty of a sinusoidal fit (solid). The shaded bands correspond to one standard deviation of phase uncertainty. b) Contrast versus momentum separation (and interferometer duration) for an $N~\hbar k$ interferometer (green dots). The dashed green line is a model of the expected contrast decay from excited state lifetime loss and the measured $\pi$-pulse efficiency, with no free parameters. The shaded band represents one standard deviation of model uncertainty. For comparison, the gray dots show the contrast of a $1~\hbar k$ interferometer with total duration equal to the associated $N~\hbar k$ sequence, and the dashed gray line is the expected contrast decay due to lifetime losses alone. Reducing the size of the atom ensemble allows for higher momentum separation (purple dots), with resolvable contrast up to $141~\hbar k$ (inset).}
     \label{fig:2}
\end{figure*} 

The LMT clock atom interferometry sequences are structured as follows (see Fig.~\hyperref[fig:1]{1c}). After an initial $\pi/2$ beamsplitter pulse, successive $\pi$ pulses are applied from alternating directions, using one interferometry beam at a time~\cite{Graham2013}. Because of the high Rabi frequency, each $\pi$ pulse interacts with both interferometer arms, toggling the atomic state in each arm and increasing the momentum separation by a net $2~\hbar k$. After the maximum momentum separation is reached, a second set of alternating $\pi$ pulses reverses the relative velocity between the arms. A third set of $\pi$ pulses decelerates the atoms such that a final $\pi/2$ pulse can close the interferometer. Thus, an $N~\hbar k$ interferometer consists of $(2N\!-\!1)$ $\pi$ pulses, where $N$ is the LMT order. We use a pulse spacing of $275~\text{ns}$, limited by the pulse generator.\smallskip

We realize Mach-Zehnder interferometers with a momentum separation varying from $1~\hbar k$ to $141~\hbar k$, without added interrogation time $T$. The overall phase of the interferometer signal can be scanned by independently adjusting the phase of the first beamsplitter pulse, leading to a sinusoidal response of the normalized excited state population (see Fig.~\hyperref[fig:2]{2a}). To analyze the interferometer signal, we produce a histogram of the normalized excited state population by marginalizing over the phase. We then extract the contrast and the offset (center) by fitting the histogram to the expected arcsine distribution, convolved with a normal distribution to include offset and amplitude noise~\footnote{The arcsine distribution has the functional form $ f(x) = n/\sqrt{1 - \left(\frac{ P_0 - x }{C/2}\right)^2}, $ with normalization factor $n$, offset $P_0$ and contrast $C$.}. Using these fit parameters as constraints, the complete interferometer signal is then fit to a sinusoid to determine the interferometer phase and its uncertainty. We find that the observed phase noise increases monotonically with the number of pulses, increasing approximately as $\sqrt{N}$ with an rms phase noise per pulse of $(50\pm 2)~\text{mrad}$. Because of the short duration of these proof-of-concept interferometers, this noise is likely dominated by intrinsic laser phase noise rather than inertial effects or vibrations.

We observe a decay in contrast with increasing momentum separation that is consistent with a combination of excited state lifetime losses and the measured $\pi$-pulse efficiency (see Fig.~\hyperref[fig:2]{2b}). To illustrate the contribution of lifetime losses alone, we study $1~\hbar k$ interferometers with variable interrogation time $T$ such that the total duration matches those of the LMT interferometers. Contrast decay due to the limited pulse efficiency likely stems from laser intensity inhomogeneity over the size of the atom ensemble. In fact, we observe increased contrast at higher LMT orders when we reduce the rms radius of the ensemble to $130~\upmu \text{m}$ (see inset).

\begin{figure}[t]
        \includegraphics[width=\linewidth]{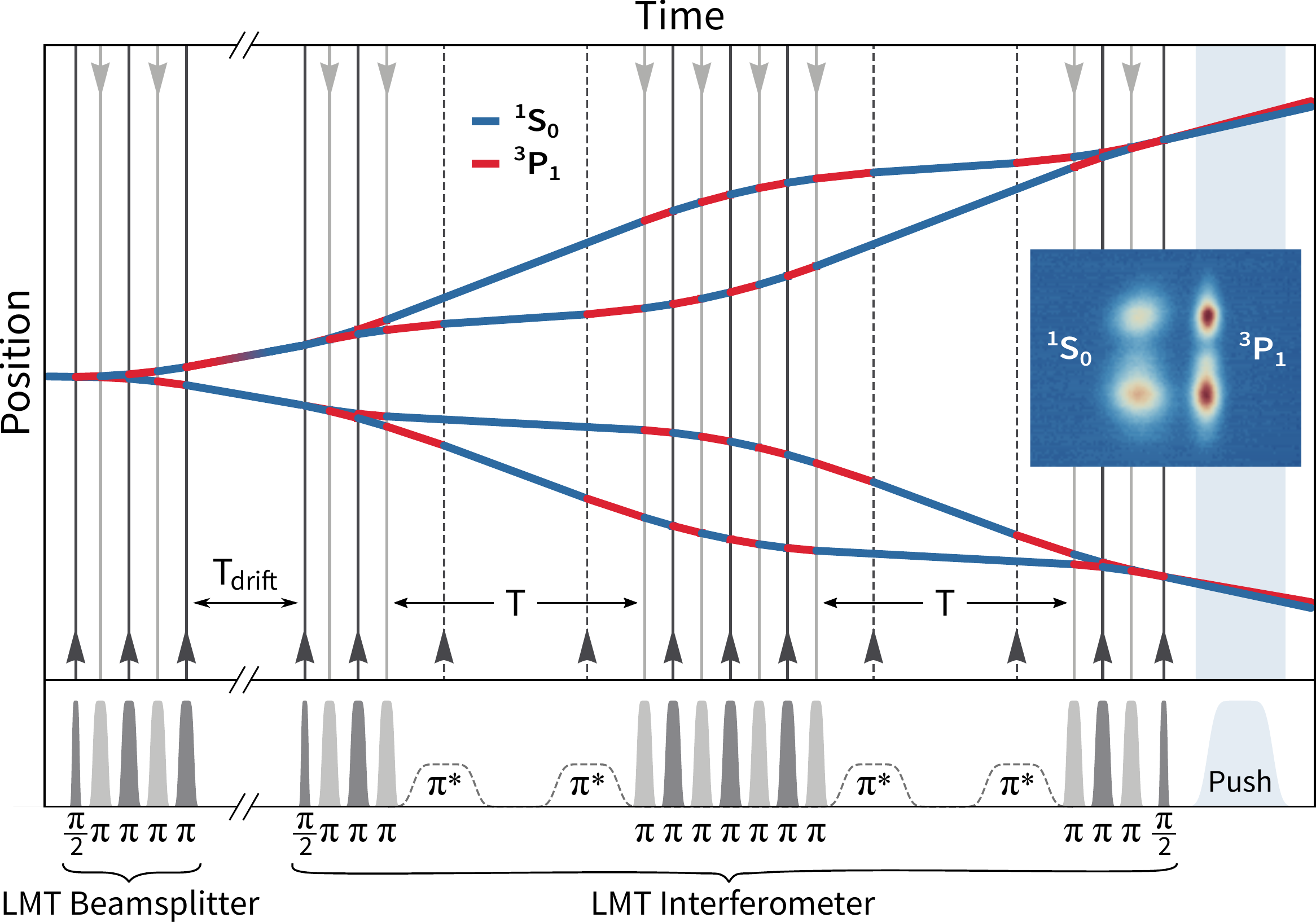}
    \caption{Example LMT gradiometer space-time diagram (top) and pulse sequence (bottom). An LMT beamsplitter and an LMT Mach-Zehnder interferometer are separated by a time $T_{\textrm{drift}}$ many times the excited state (red) lifetime, such that both the upper and the lower interferometer start off with all atoms in the ground state (blue). Despite the large relative velocity of the interferometers, every pulse interacts with all interferometer arms due to the high Rabi frequency. Individual arms can be addressed using longer, lower intensity pulses (dashed, $\pi^*$) with reduced Doppler bandwidth. The excited state population can then be stored in the ground state during the interrogation time $T$ to avoid spontaneous emission loss. An example fluorescence image of the upper and lower interferometer ports is shown as an inset.
    }\label{fig:3}
\end{figure}

\begin{figure}[t]
        \includegraphics[width=\linewidth]{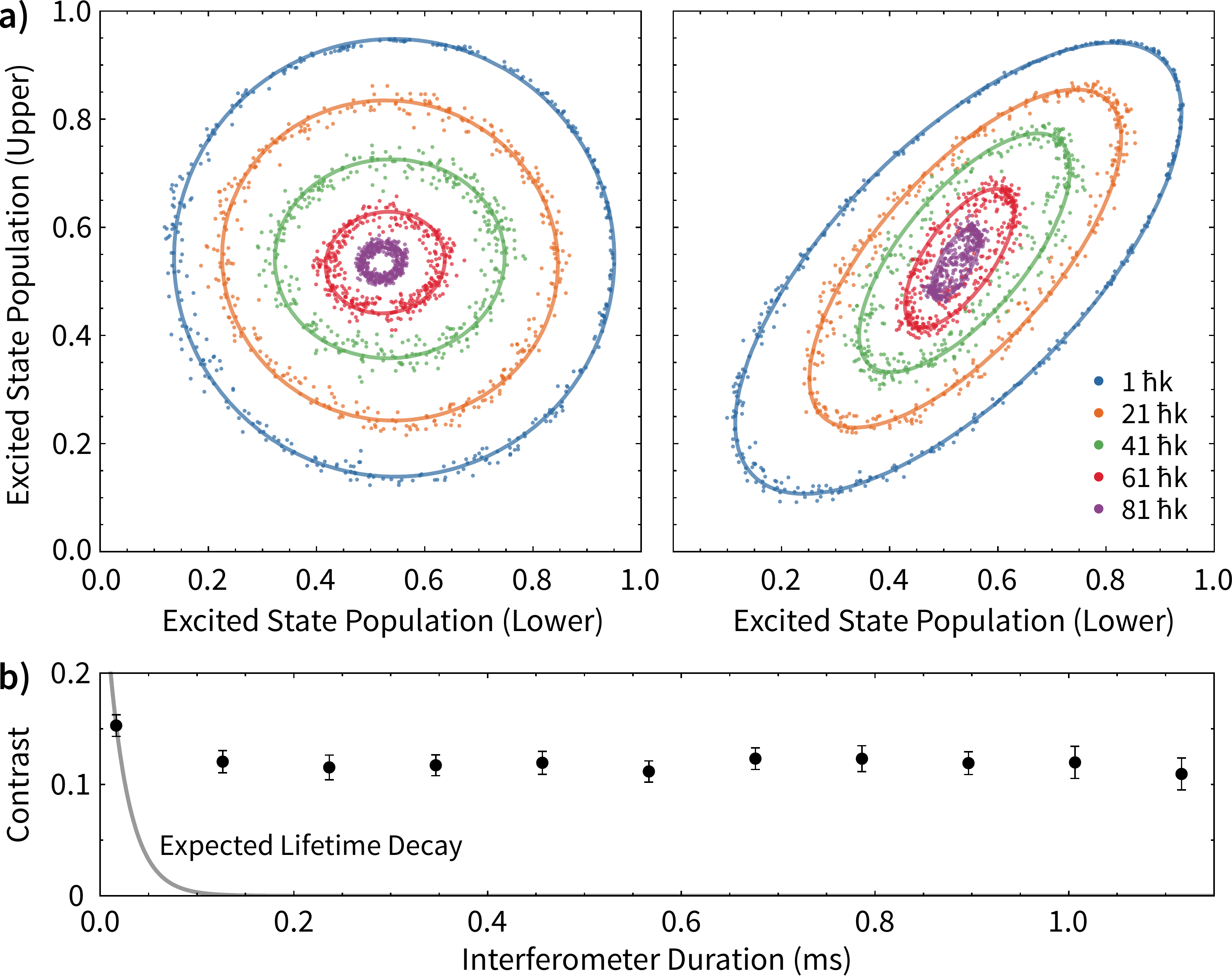}
    \caption{a) Normalized excited state populations for LMT-enhanced gradiometers from $1 \hbar k$ to $81 \hbar k$, at an applied differential phase of approximately {$90$\textdegree} (left) and {$45$\textdegree} (right). The gradiometer contrast is reduced at larger momentum separation, consistent with the LMT interferometer results (see Fig.~\hyperref[fig:2]{2b}). b) Gradiometer contrast versus interferometer duration, using velocity-selective pulses to store the excited state population in the ground state during the interrogation time. The solid line represents the expected lifetime decay without velocity-selective pulses. We extend the interferometer duration to over 50 times the $21.6~\upmu \text{s}$ excited state lifetime without any additional loss of contrast.}
    \label{fig:4}
\end{figure}

To demonstrate how laser phase noise in the interferometers can be suppressed in a differential measurement, we realize LMT-enhanced clock gradiometers by splitting the atom ensemble with an LMT beamsplitter prior to the interferometry sequence (see Fig.~\hyperref[fig:3]{3}). A time delay $T_\text{drift}$ ensures that all excited state atoms have decayed so that both interferometers start off in the ground state. The relative velocity $\Delta v = N_{\text{BS}}~ \hbar k/m$ between the interferometers gives rise to a differential Doppler shift, where $N_{\text{BS}}$ is the beamsplitter LMT order. Because of the high Rabi frequency, each pulse still interacts with both arms of both interferometers. While the individual interferometers are subject to laser phase noise, their phases are highly correlated and the normalized populations trace out an ellipse (see Fig.~\hyperref[fig:4]{4a}). We control the differential phase in the gradiometer by delaying the final $\pi/2$ pulse by $\Delta T$. This leads to a relative phase shift between the interferometers of $\Delta \phi = \omega_a~ \Delta T~\Delta v/c$, where $\hbar \omega_a$ is the atomic energy splitting of the transition~\cite{Graham2013}. We implement LMT gradiometers with a relative velocity corresponding to $N_{\text{BS}} = 51$ and vary the momentum separation of the interferometer from $1~\hbar k$ to $81~\hbar k$, for a selection of differential phases $\Delta \phi$ and without added interrogation time $T$.

The differential phase is the output signal of the gradiometer measurement. Its extraction is limited by three primary noise sources: differential phase noise, amplitude noise, and offset noise~\cite{Stockton2007}. We analyze each ellipse using maximum likelihood estimation (MLE) with a model including offset noise for each individual interferometer as well as differential phase noise. We find that our LMT-enhanced gradiometers are dominated by offset noise with an average magnitude of $1\%$, leading to homogeneous broadening of the ellipse.
We infer the differential phase with an average uncertainty of $60~\text{mrad}$ per shot, limited primarily by this offset noise. This bounds our ability to measure any residual differential phase noise to the same level. The offset noise likely stems from residual atoms in unresolved, neighboring velocity classes in each interferometer port as a result of the finite pulse efficiency.

The LMT gradiometer features the same contrast decay observed in the individual Mach-Zehnder interferometers (see Fig.~\hyperref[fig:2]{2b}). However, the total duration of the gradiometer can be extended far beyond the excited state lifetime by storing the atoms in the ground state during the interrogation time. This requires selectively inducing transitions in only one arm of each interferometer at a time, which we accomplish with velocity-selective pulses using a lower Rabi frequency (see Fig.~\hyperref[fig:3]{3}). The duration of these pulses is carefully chosen to act as a $\pi$ pulse for one arm and a $2\pi$ pulse for the other~\footnote{For an $N \hbar k$ interferometer, this condition is satisfied at a Rabi frequency $\Omega = \frac{\Delta}{\sqrt{3}} = \frac{1}{\sqrt{3}} N \frac{\hbar k^2}{m}$}. To address both interferometers simultaneously and maintain common-mode laser noise suppression, the velocity-selective pulses are generated with two separate RF signals applied to the same AOM, with the frequencies separated by the relative Doppler shift of the interferometers. We use a relative velocity corresponding to $N_{\text{BS}} = 81$ and an interferometer momentum separation of $31~\hbar k$, with a Rabi frequency of $500~\text{kHz}$ for the velocity-selective pulses. With these parameters, we extend the total interferometer duration to $1.12~\text{ms}$ ($T = 0.55~\text{ms}$) without any additional loss of contrast in the gradiometer (see Fig.~\hyperref[fig:4]{4b}). In our setup, the interrogation time is limited by the atoms falling out of the horizontal interferometry beams, which can be avoided with a vertical beam geometry.

While we demonstrate how to circumvent interrogation time limitations posed by the excited state lifetime, further extending the LMT order requires reducing the contrast decay. Imperfect pulse efficiency due to inhomogeneous broadening can be suppressed with improved spatial filtering and by increasing the laser beam diameter. Losses due to the finite excited state lifetime can be minimized by using shorter pulse durations. Both of these technical limitations can be mitigated by using more laser power. We estimate that increasing the power per beam from $100~\text{mW}$ to $3~\text{W}$ would enable a $1000~\hbar k$ interferometer at approximately $10\%$ contrast. Moreover, this work serves as a proof of principle for future LMT-enhanced clock atom interferometry on narrower spectral lines such as the ${}^1S_0~ \mhyphen {}^3P_0$ clock transition in \textsuperscript{87}Sr, where lifetime losses can be negligible and the resulting pulse efficiency can support many thousands of consecutive pulses. The long lifetime is required to mitigate spontaneous emission loss due to the light propagation delay over long baselines. Therefore, such a transition can address the ambitious LMT requirements for gravitational wave detection and dark matter searches with atomic sensors~\cite{Graham2017}. Finally, an alternative implementation could employ both spectral lines in a two-color clock atom interferometer, where the ${}^1S_0~ \mhyphen {}^3P_1$ transition is used for fast and efficient momentum transfer, and the ${}^1S_0~ \mhyphen {}^3P_0$ transition for velocity-selective pulses and extended interrogation times.

Although our interferometry laser is frequency stabilized to an optical cavity, this is not generally required since neither the linewidth of the laser nor the temperature of the atoms affect the pulse efficiency in the high Rabi frequency limit. The resulting Doppler insensitivity makes clock atom interferometry on the $689~\text{nm}$ transition promising for applications in gravimetry and mobile inertial sensing. For instance, an accelerometer with sensitivity below $10^{-9}g/\sqrt{\text{Hz}}$ ($1~\text{$\upmu$Gal}/\sqrt{\text{Hz}}$) can be realized with $100~\hbar k$ atom optics, $1~\text{mrad}/\sqrt{\text{Hz}}$ read noise~\footnote{To reduce read noise to this level, laser phase noise can be suppressed as a common-mode by deriving sequential counter-propagating pulses from a single laser pulse using a delay line.}, and $T=10~\text{ms}$ interrogation time, allowing for a repetition rate of over $10~\text{Hz}$. Such a sensor could be implemented using a broadband strontium red MOT at a temperature of around $100~\upmu\text{K}$~\cite{Loftus2004} without the need for cavity linewidth reduction and would only require minimal magnetic shielding compared to alkali atoms~\cite{Ferrari2003,delAguila2018} commonly used in mobile atomic sensors~\cite{Biedermann2015,Farah2014,Rudolph2015,Freier2016}.

\medskip

This work was supported by the Office of Naval Research Grant No.~N00014-17-1-2247 and by the U.S. Department of Energy, Office of Basic Energy Sciences under Award No.~DE-SC0019174. T.W.\ is supported by the National Defense Science and Engineering Graduate Fellowship. We thank Mark Kasevich for helpful discussions.

\end{document}